\documentclass[10pt,twocolumn]{article}
\usepackage[letterpaper,margin=0.67in,columnsep=0.24in]{geometry}
\usepackage[T1]{fontenc}
\usepackage{tgtermes}
\usepackage[scale=0.95]{tgheros}
\usepackage{microtype}
\usepackage{graphicx}
\usepackage{booktabs}
\usepackage{tabularx}
\usepackage{enumitem}
\usepackage{xcolor}
\usepackage{hyperref}
\usepackage{caption}
\usepackage{authblk}
\usepackage[authordate,backend=biber,maxcitenames=2,minbibnames=3,maxbibnames=4]{biblatex-chicago}
\definecolor{ink}{HTML}{172033}
\definecolor{accent}{HTML}{8B1E3F}
\hypersetup{colorlinks=true,linkcolor=accent,citecolor=accent,urlcolor=accent,pdfauthor={Fred Zimmerman},pdftitle={Copyright Is the Headline; Capability Is the Blind Spot},pdfsubject={AI coverage in the book-publishing trade press, November 2025--August 2026}}
\setlist{nosep,leftmargin=1.15em}

\newcommand{\bottomline}[1]{\noindent\colorbox{accent!8}{\parbox{0.94\columnwidth}{\small #1}}}

\title{\vspace{-1.6em}\textbf{Copyright Is the Headline; Capability Is the Blind Spot}\\
\large AI Technology in the Book-Publishing Trade Press, November 2025--August 2026}
\author{Fred Zimmerman}
\affil{Nimble Books LLC, Ann Arbor, Michigan\\\texttt{wfz@nimblebooks.com}}
\date{August 1, 2026}

\begin{document}
\twocolumn[
\maketitle
\begin{abstract}
This rapid evidence review examines 89 articles about artificial intelligence (AI) and book publishing published from November 1, 2025 through August 1, 2026. The purposive corpus spans English-, Chinese-, German-, French-, Spanish-, Portuguese-, Italian-, and Japanese-language publishing coverage; major-newspaper book coverage; and specialist technology commentators. Each item was coded for topic, stance, technical depth, and dominant voice. The press is neither silent nor simply hostile: 30\% of items are risk-framed, 42\% mixed, and 28\% opportunity-framed. Chinese coverage is markedly operational and opportunity-oriented; specialist commentary is substantially deeper than trade reporting. Yet the corpus still clusters around rights, licensing, governance, reader trust, workflow adoption, and product announcements. Only ten items offer sustained technical scrutiny, and none centers a direct interview with a frontier-lab researcher or evaluation engineer. The central gap is reporting that connects model architecture and evaluation to publishing decisions: capability elicitation, RAG, prompt injection, agent reliability, inference economics, model drift, provenance, reader research, and reproducible workflow evaluation. The report recommends a standing AI beat, recurring lab interviews, a claims ledger, shared test protocols, reader panels, and technical columnists. The supplement supplies the full coded corpus, claim ledger, glossary, and BibTeX database.
\end{abstract}
\vspace{1.2em}
]

\section*{Bottom Line Up Front}
\bottomline{The trade press recognizes the stakes but still reports AI mainly as lawsuits, scandals, policies, and launches. That identifies who is fighting and what was announced; it rarely establishes what a system can do, under which conditions, at what cost, with which failure modes, or for how long the answer remains valid. The missing link is technical adversarialism: interview the labs, inspect the evaluation harness, reproduce the workflow, disclose the economics, and test the reader.}

\section{Motivation}

The review begins from five observed disconnects. First, publishing debates ``AI'' as if it were one stable technology, while deployed systems differ materially in tool access, context management, retrieval, reasoning budget, safeguards, and post-training. Second, a legal vocabulary---training, fair use, licensing, consent---often substitutes for a technical account of what happens after a model is trained or connected to a catalog. Third, the people most affected (authors, translators, narrators, readers, editors) are visible, but the people who build and evaluate frontier systems are largely absent. Fourth, claims about reader rejection or acceptance outrun the limited and contradictory evidence. Fifth, the study period encompasses a dramatic step change in the efficacy of agentic AI that began in late 2025. Laboratory releases paired stronger models with context compaction, tool use, and support for longer-running agents; an independent task-horizon series documents the broader rise in the duration of software tasks frontier agents complete reliably. The laboratory capability claims are self-reported; the trend evidence is independently measured \parencite{ANTHROPIC2025OPUS45,OPENAI2025GPT52,METR2026HORIZONS}.

The stakes are high. Publishing allocates cultural attention, certifies knowledge, and converts rights into long-lived assets. AI can lower production and localization costs, but also makes industrial-scale imitation, fraud, and market flooding cheaper. The pace is unusually fast: within this nine-month window, the corpus moves from voice-AI conference programming and marketing pilots to a major-house cancellation over suspected machine authorship, publisher suits against multiple model providers, large-scale platform rejection of low-quality submissions, and interactive books that blur reading with software use \parencite{C01,C02,C24,C44,C33}.

The moment is also one of controversy and stock-taking. UK policy moved away from a preferred broad training exception; EU transparency duties approach enforcement; US litigation continues to distinguish model training from the acquisition of pirated copies; and Japan's copyright framework remains materially different \parencite{UKGOV2026,EUCOM2026,BUNKA2024,C49}. A useful trade press must keep these jurisdictions separate while connecting them to the same operational question: what should a publisher do on Monday morning?

\section{Corpus and Method}

The unit of analysis is an article, essay, or news report primarily concerned with AI's effect on book publishing. The date window is November 1, 2025 through August 1, 2026. Discovery combined outlet archive pages, topic pages, multilingual web search, citation chaining, commentator archives, and targeted searches for named incidents and institutions. The final purposive corpus contains 89 items: 61 trade-journal pieces, 14 major-newspaper pieces, and 14 specialist-commentator essays. It includes 31 US-coded items, 16 UK, ten China, seven international, four each from France, Spain, Brazil, and Italy, three Germany, two Japan, and four continental-EU items not assigned to a single country.

This is an analytic sample, not a census. Paywalls, weak indexing, inconsistent tags, and language access affect inclusion. US and UK coverage remains overrepresented. The expanded corpus includes Chinese, German, French, Spanish, Portuguese, Italian, and Japanese publishing sources. Non-English claims are analytic translations; titles retain the source language where practicable. The commentator stratum includes Thad McIlroy, Jane Friedman, Kathleen Schmidt, Josh Bernoff, Dosdoce, Actualidad Editorial, and PublishNews columnists. Association and government documents test journalistic claims but are not counted as coverage unless published as journalism or commentary.

Each item received one primary and one secondary category from an open-ended codebook: legal/regulatory; business/licensing; production/workflow; discovery/marketing; reader experience; provenance; fraud/security; governance; translation; market structure; ethics/culture; technology; and data governance. Stance is risk, mixed, or opportunity. Technical depth runs from 0 (``AI'' is only a label) to 3 (mechanism, evidence, boundary conditions, and failure modes receive scrutiny). The ``dominant voice'' records whose account structures the article. Coding was performed once by the author; no intercoder reliability is claimed. All item-level codes and claims are supplied in the supplement and CSV.

\begin{table}[t]
\centering
\caption{Top primary topics in the coded corpus (n=89).}
\small
\begin{tabular}{lrr}
\toprule
Topic & $n$ & \% \\
\midrule
Business/licensing & 13 & 14.6 \\
Governance & 13 & 14.6 \\
Reader experience & 12 & 13.5 \\
Production/workflow & 10 & 11.2 \\
Legal/regulatory & 10 & 11.2 \\
Market structure & 7 & 7.9 \\
All other topics & 24 & 27.0 \\
\bottomrule
\end{tabular}
\end{table}

\section{What the Trades Say}

\subsection{The legal story became a business story}

Legal/regulatory coverage is tied with reader experience as the largest primary category. Articles trace suits over training data, shadow libraries, voice and identity imitation, platform liability, and the Anthropic settlement \parencite{C21,C25,C29,C34,C47,C49}. The best pieces preserve an essential distinction often lost in public debate: a finding that a use in training may be fair does not legalize acquisition from pirate sites. The settlement resolves claims about pirated copies while leaving the training holding intact \parencite{C49}. Jurisdiction-aware reporting also improved. UK coverage followed the government's retreat from a preferred opt-out exception, while European reporting discussed collective enforcement and remuneration \parencite{C23,C14}.

Coverage then pivots toward a licensing market. Cashmere is presented as infrastructure for licensing, monitoring, and monetizing content in AI systems; London Book Fair coverage treats TDM and RAG licenses as active markets; Ingram lets publishers refuse sales to technology firms suspected of scanning print books \parencite{C16,C19,C10}. McIlroy supplies the missing distinction: irreversible pretraining and attributable, revocable RAG access are different products with different economics \parencite{C53}. Bernoff, meanwhile, stress-tests Ingram's opt-out and finds obvious secondhand and intermediary routes around it \parencite{C58}. Commentators are doing technical due diligence that reported announcements often omit.

\subsection{Provenance became a crisis of process}

The \emph{Shy Girl} episode dominates late-March coverage. The New York Times reported the cancellation using detector scores and textual signals; Publishers Weekly and the Guardian widened the frame to disputed facts, asymmetric author-publisher rules, due process, and detector fallibility \parencite{C24,C28,C30}. Jane Friedman extends the problem to prizes: a prohibition without definitions, evidence thresholds, and appeal procedures can punish innocent writers while failing to deter sophisticated use \parencite{C38}. This is the corpus at its most conceptually mature. It begins to replace the binary ``human or AI?'' with a chain-of-custody question: who did what, to which version, with which tool, under what agreement?

Yet most proposed remedies remain disclosure forms and detectors. Independent research supports caution. Current detectors vary by genre, model, editing, and language; hybrid text is especially difficult, and evasion changes the target \parencite{DOI101007s4097902600226w,DOI101016jcompedu2026105616}. Cryptographic provenance standards can record origin and transformations, but they cannot prove that an uncredentialed text is synthetic or that every declared step is truthful \parencite{C2PA2026}. The appropriate analogy is food traceability, not a lie detector.

\subsection{Operational AI is visible but under-tested}

The opportunity articles cover audiobook production, interactive editions, catalog marketing, manuscript positioning, microdrama adaptation, metadata, inventory forecasting, and routine workflow automation \parencite{C04,C06,C08,C12,C15,C31,C41,C52}. The most technically grounded trade item in the sample reports both uses and limits: cheaper production does not answer what to publish, for whom, or how to form an audience; identical inputs can yield undifferentiated strategies; trend analysis is not prediction \parencite{C52}. The Frankfurt marketing panel is similarly valuable because its on-stage AI agent failed and prompted discussion of confirmation bias \parencite{C02}.

But vendor claims usually arrive without baselines. ``Faster,'' ``scalable,'' and ``data-driven'' are rarely tied to an error taxonomy, held-out test set, human-review time, rights status, cost per accepted asset, or downstream sales lift. In an industry that routinely asks for comparable-title evidence, its technology reporting too often accepts the technological equivalent of jacket copy.

\begin{figure*}[t]
\centering
\includegraphics[width=0.98\textwidth]{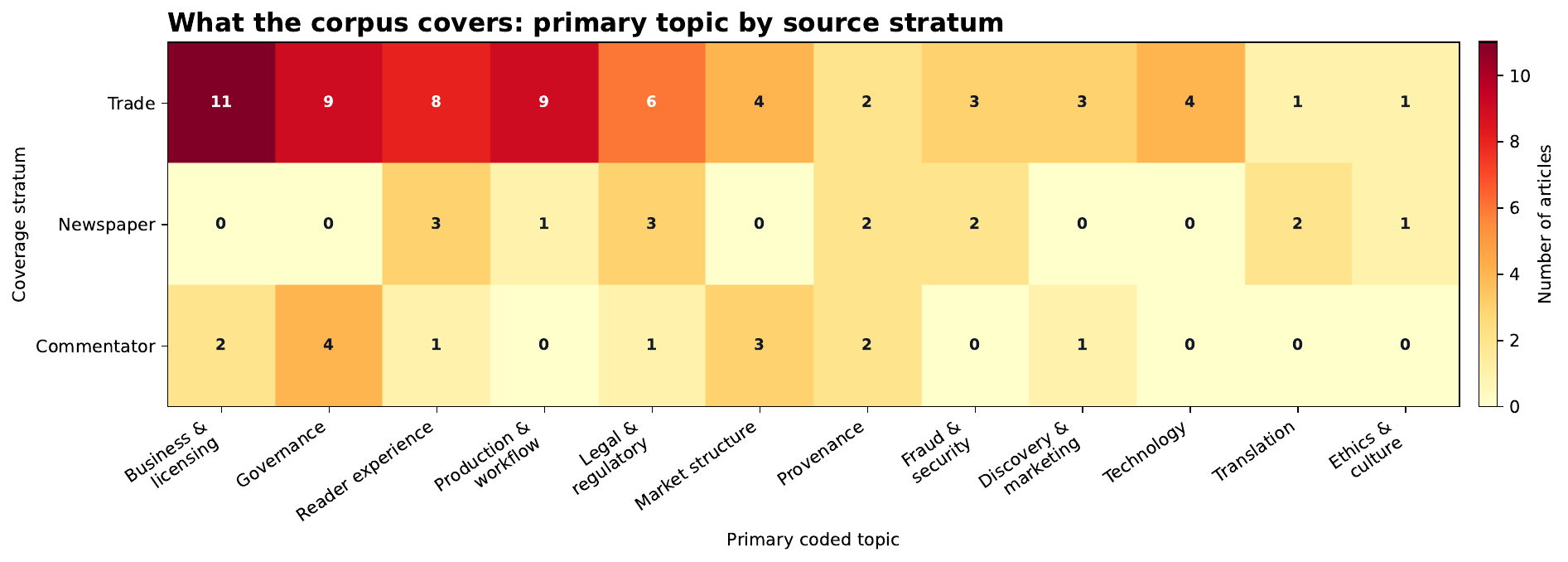}
\caption{Primary topic by source stratum. Trade coverage is broad; business/licensing leads, with governance, workflow, and reader experience close behind. Categories were assigned once by the author; counts are descriptive, not population estimates.}
\end{figure*}

\subsection{Readers appear as a claim more often than a population}

Waterstones would stock clearly labeled AI books if readers wanted them, but recoils from the prospect; interactive-book coverage imagines reader value while warning about reliability; commentary on \emph{Shy Girl} argues that suspicion itself can warp reading \parencite{C05,C08,C27}. These are plausible positions, not a market map.

The external evidence is unsettled. A Swedish Kantar survey ($n=1{,}031$) finds strong stated resistance to AI-generated books \parencite{SFF2026}. Experimental studies, by contrast, show that readers may fail to distinguish origins, that disclosure can depress evaluations, and that model outputs can sometimes match or outperform human text in constrained tasks \parencite{KIM2025,MARKLOVA2026,MATULIC2026,DOI1010801050842220262696829}. These results are not contradictory: stated purchase ethics, blind aesthetic judgment, long-form satisfaction, and willingness to pay are different outcomes. The trades seldom separate them.

\emph{Daggermouth} makes that evidentiary problem concrete. The viral novel became a bestseller and attracted a reported seven-figure traditional deal after a research dataset flagged a large share of its text as likely AI-generated; H. M. Wolfe denied using generative AI. The case matters not as proof of machine authorship---the public evidence does not establish that---but because commercial success and reader enthusiasm coexist with disputed detector evidence. It therefore breaks the easy equation ``AI-written equals unsellable slop'' while strengthening the case for upstream provenance and due process \parencite{C51,CHAKRABARTY2026}.

\subsection{Multilingual coverage reveals unequal effects}

The strongest international item examines translation data and a small round-trip experiment. It argues that automation may widen the advantage of high-resource languages, handle technical sentences better than literary passages, and shift work toward expert post-editing \parencite{C32}. Spanish coverage adds translators' labor and a proposed human-translation label \parencite{C80}; French coverage follows consumer deception and title-level licensing \parencite{C74,C75}. German and Italian trades contribute evidence absent from most Anglophone reports: adoption surveys, an operational publisher RAG system, explicit agent training, and disclosure taxonomies that reject a binary AI/no-AI seal \parencite{C70,C71,C72,C85,C87}. Portuguese-language coverage brings Brazilian governance, catalog strategy, and workflow training into view \parencite{C82,C83,C84}.

The Japanese stratum remains thin, but it is no longer purely cultural. Murakami's claim that his literature is distinct from AI literature registers resistance, while the Japan Electronic Publishing Association explicitly connects scaling, multimodality, reasoning models, and agents to publishing work \parencite{C46,C89}. Major gaps remain: Korean, Arabic, Hindi and other South Asian languages, African publishing markets, manga/comics, vertical writing, and local rights metadata receive little or no direct coverage in the sample.

\subsection{Chinese coverage changes the comparative picture}

The ten Chinese items are not a token regional add-on; they change the frame. Seven are opportunity-led and none is risk-led. Four center reader experience, three production workflow, two business/licensing, and one technology. Coverage emphasizes proprietary platforms, domain knowledge graphs, AI reading companions, full-workforce training, and movement from discrete books toward continuous knowledge services \parencite{C54,C55,C56,C60,C61,C69}. Even governance is often presented as infrastructure: 22 organizations propose a standardized, rights-cleared, commercially usable corpus \parencite{C63}.

That does not make Chinese coverage naively boosterish. Articles defend deep reading, human developmental judgment in children's books, and reader-centered evaluation \parencite{C65,C67,C68}. But the center of gravity differs sharply. In this purposive sample, Chinese reporting asks how institutions can build and govern AI-enabled publishing capacity; Anglophone coverage more often asks who was harmed, copied, deceived, or sued. The comparison exposes a blind spot on both sides: operational reporting can mute conflict and labor, while conflict reporting can miss infrastructure and implementation.

\subsection{Commentators supply depth---and strong priors}

The 14 specialist-commentator items average 2.14 on the 0--3 depth scale, versus 1.54 for the 61 trade items. McIlroy distinguishes training from retrieval and relates AI disruption to the industry's earlier desktop-publishing transition \parencite{C53,C62}. Bernoff brings an explicit ethical code, adversarial analysis of opt-outs, and a 1,481-writer survey rather than impressions \parencite{C57,C58,C64}. Friedman follows governance and generative-engine discoverability; Schmidt foregrounds deception, market flooding, and reader trust \parencite{C38,C43,C66,C59}. Spanish and Brazilian commentators extend the analysis to backlists as datasets, local discoverability, and AI as platform rather than isolated tool \parencite{C79,C82}.

Their value is precisely that they argue. Their limitation is the same: argument is not measurement, and strong priors can harden into untested assumptions. The widened corpus therefore strengthens the case for named technical columnists while also strengthening the case for shared tests, disclosed evidence, and separation of analysis from advocacy.

\section{Gap Analysis}

Figure~2 shows the structural problem. Technical researchers dominate only five of 89 articles; no item is structured around a direct interview with an OpenAI, Anthropic, or Google DeepMind researcher or evaluation engineer. Platforms and vendors are present, but product spokespeople are not substitutes for people who can explain training, post-training, evaluation, retrieval, safeguards, and failure analysis.

\begin{figure*}[t]
\centering
\includegraphics[width=0.98\textwidth]{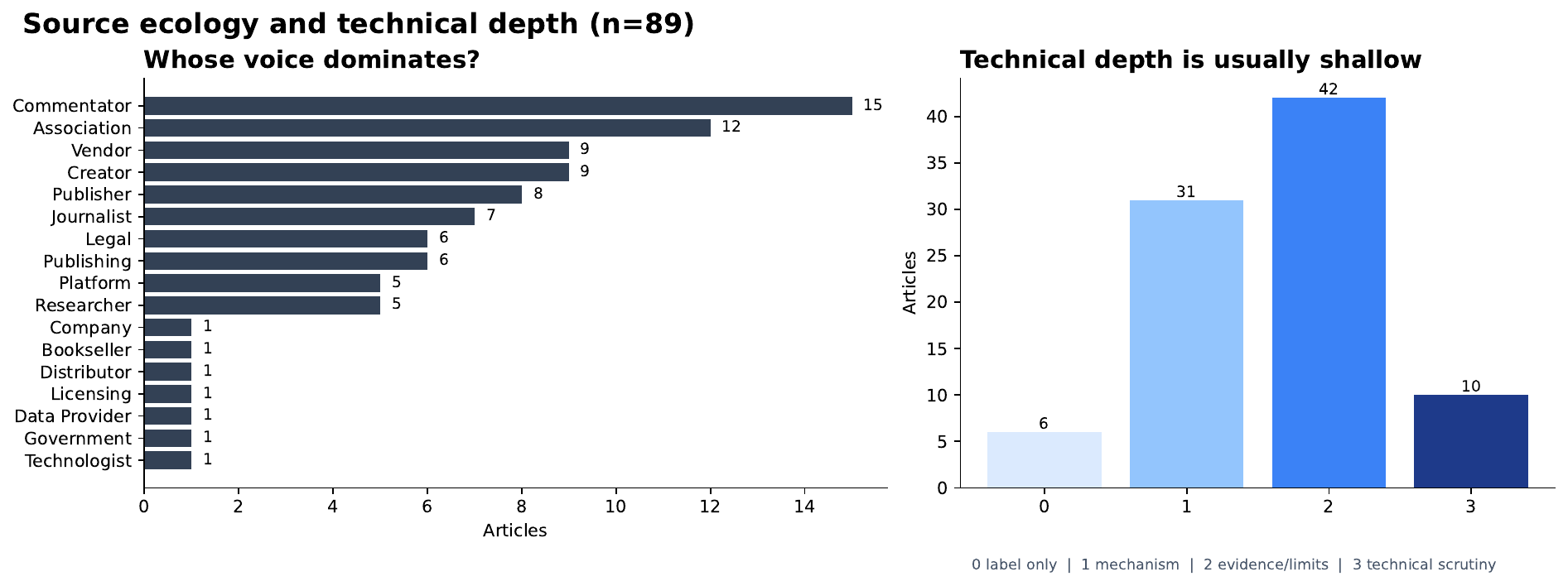}
\caption{Dominant voice and technical-depth score. Ten articles reach depth 3; the mean is 1.63 on a 0--3 scale. Commentators average 2.14 and trade items 1.54. No sampled article centers a frontier-lab researcher or evaluation engineer.}
\end{figure*}

\subsection{Seven missing beats}

\textbf{1. Capability is conditional.} A model does not have one stable ``IQ.'' Results depend on instructions, reasoning budget, tools, retrieval, memory, retries, and the surrounding harness. \emph{Elicitation} is the work of drawing out the strongest credible performance. A weak prompt can understate capability; an overfit or leaky benchmark can exaggerate it \parencite{HOFSTATTER2025,OPENAI2026ELICIT}. Trade claims should name the full tested system, not only a brand.

\textbf{2. Evals are publishing's missing instrument panel.} Frontier labs distinguish general capability tests from task-specific evaluations. A publisher needs tests for citation fidelity, editorial error detection, voice consistency, metadata completeness, accessibility, translation adequacy, and rights compliance, with predeclared pass thresholds and human baselines \parencite{OPENAI2025EVALS,OPENAI2026DEPLOY}. Almost no trade article reports such a test.

\textbf{3. RAG is not licensing shorthand.} Retrieval-augmented generation supplies selected documents at answer time rather than baking them into model weights. It enables metered access and citation, but introduces retrieval failure, access-control, and prompt-injection risks. A malicious or merely malformed document can manipulate a downstream agent; policy enforcement should occur before sensitive text reaches the generator \parencite{SDRAG2026,RAGSEC2026}. The corpus praises RAG licensing without covering this attack surface.

\textbf{4. Agents change workflow risk.} A chatbot drafts; an agent can search, call tools, modify files, and publish. Reliability compounds across steps, and logs, permissions, rollback, and human approval become editorial controls. Coverage of workflow automation rarely distinguishes assistance from delegated action.

\textbf{5. Economics require unit measures.} Vendor funding and valuations are news, but publishers need cost per accepted page, localized title, verified citation, qualified lead, and incremental sale. Include compute, integration, review, correction, indemnity, and model migration. Without a denominator, ``15\% faster'' can be economically meaningless.

\textbf{6. Reader research needs factorial design.} Measure generation share, disclosure wording, genre, author reputation, price, sample length, and human editing separately. Record both blind quality and informed willingness to buy. Do not generalize from short passages to novels or from moral attitudes to behavior.

\textbf{7. Provenance must move upstream.} Detectors infer from style after the fact. Publishers should preserve version history, contributor attestations, tool logs, source permissions, and transformation records from acquisition onward. Standards such as C2PA suggest a signed-manifest architecture, but contracts and production systems must make it routine \parencite{C2PA2026}.

\section{Filling the Gaps}

\subsection{Recommendations for trade editors}

\begin{enumerate}
\item \textbf{Create a standing Future of the Book interview.} Monthly, interview one researcher or engineer from Google DeepMind, OpenAI, Anthropic, Meta, Microsoft, a leading open-source lab, or an independent evaluator. Ask the same questions about system boundaries, elicitation, eval design, failure modes, costs, rights, privacy, and six-month expectations. Pair the lab voice with an author, publisher, librarian, or translator.
\item \textbf{Appoint two complementary columnists.} One should understand publishing rights and labor; the other should be able to inspect code, system cards, model evaluations, security, and data. Rotate regional correspondents, especially Japan, China, Korea, India, Latin America, Africa, and smaller-language Europe.
\item \textbf{Publish a claims ledger.} For every product or policy claim, record claimant, evidence, baseline, jurisdiction, affected workflow, measurement date, uncertainty, and what would falsify it. Update rather than overwrite as models and cases change.
\item \textbf{Build a small test kitchen.} Maintain a rights-cleared benchmark pack: manuscripts, ONIX, covers, contracts, translations, audio passages, and accessibility tasks. Re-run leading systems quarterly with fixed rubrics and publish prompts, settings, human-review time, errors, and cost.
\item \textbf{Convene a reader panel.} Combine a longitudinal consumer panel with preregistered blind tests. Segment professional readers, genre communities, accessibility users, and general buyers. Report behavior as well as attitudes.
\item \textbf{Adopt incident-reporting norms.} When AI is implicated in a cancellation, fraud, leak, or faulty release, report the version chain, evidence standard, affected rights, remediation, appeal path, and lessons. Avoid detector percentages without uncertainty and corroboration.
\end{enumerate}

\subsection{A reading list for the beat}

Start with the UK government's 2026 copyright report, EU transparency guidance, and Japan's 2024 general understanding for jurisdictional grounding \parencite{UKGOV2026,EUCOM2026,BUNKA2024}. Add the BISG/BookNet survey for industry practice; recent publishing-specific research on adoption and forecasting; detector reliability studies; reader-disclosure experiments; OpenAI's business-evaluation primer and third-party evaluation playbook; and current RAG security work \parencite{BISG2026,DOI103389frma20261759242,DOI101007s1210902610070y,DOI101007s4097902600226w,MATULIC2026,OPENAI2025EVALS,OPENAI2026ELICIT,RAGSEC2026}. The point is not to canonize lab publications. It is to equip journalists to interrogate them.

\section{Limitations and Conclusion}

This review is purposive, time-bounded, unevenly multilingual, and single-coded. Search visibility, paywalls, translation, and unequal archive access affect the sample; US and UK sources still constitute more than half the corpus, while several major publishing languages are absent or represented by one item. ``Dominant voice'' simplifies multi-source articles, and technical depth reflects reporting detail rather than whether an article reached the correct policy conclusion. The results should guide a larger multilingual content analysis, not stand in for one.

Still, the signal is clear. Publishing's trade press has graduated from asking whether AI matters. It reports real litigation, labor harms, licensing ventures, platform contamination, and operational experimentation. Its next task is harder: connect those developments to how modern AI systems are actually built, evaluated, secured, priced, and used. The missing beat is not futurism. It is technical accountability. As the newsroom saying goes, if your mother says she loves you, check it out; if a vendor says its AI transforms publishing, inspect the eval.

\printbibliography[title={References}]

\end{document}